\documentclass[11pt]{article}
\begin{document}
\begin{flushright}
SJSU/TP-05-26\\
Aug 2005\end{flushright}
% Version 2
\vspace{1.7in}
\begin{center}\Large{\bf On relational time and energy decoherence}\\
\vspace{1cm}
\normalsize\ J. Finkelstein\footnote[1]{
        Participating Guest, Lawrence Berkeley National Laboratory\\
        \hspace*{\parindent}\hspace*{.5em}
        Electronic address: JLFINKELSTEIN@lbl.gov}\\
        Department of Physics\\
        San Jos\'{e} State University\\San Jos\'{e}, CA 95192, U.S.A
\end{center}
\begin{abstract}
Students Alice and Bob take an examination in their quantum mechanics
class, and thereby illustrate some aspects of energy decoherence.
\end{abstract}
\newpage
Quantum mechanics is a beautiful theory.  It is also, in the hands of
skilled practitioners, a wildly-successful theory for microscopic
phenomena, in spite of the
contradiction between the world as we experience it and the
superpositions of macroscopically-distinguishable states which (at
least naively) would seem to be implied by  unitarity of evolution in
time. 

The ``time'' in which quantum states are supposed to evolve unitarily
is a parameter, not a dynamical variable of the theory.  On the other
hand, any actual determination of time would involve the reading of a
real clock, which is a physical (and hence a quantum) system.  It has
been suggested 
(for example, in refs.~\cite{SW}--\cite{GPP}) that one should study 
the evolution of a
quantum system not with respect to an abstract time parameter, but
rather with respect to the reading of a clock considered as a physical
system. I will refer to this reading as the ``relative time''.  As I
will review below, the evolution of a quantum system with respect to a
real (and hence not infinitely-accurate) clock will involve
decoherence in energy of that system, and it has also been 
suggested (for example, in refs.~\cite{GPP}--\cite{BW}) 
that decoherence in energy could at least
alleviate the alleged conflict between quantum predictions for
microscopic phenomena and the appearance of the macroscopic world.      

A seemingly-different approach (which does not involve the explicit
consideration of clocks as physical systems) is to assume that in fact
the evolution of quantum states is not unitary, but rather
that unitary evolution is supplemented by
collapse onto states of definite total 
energy~\cite{BW}--\cite{AH}. I will refer to these theories as 
``energy-driven
collapse theories,'' and will review below the fact that in these
theories quantum states experience energy decoherence just as they do  in
relative-time calculations.

It would certainly be interesting if it could be shown that either
relative-time calculations or energy-driven collapse theories made 
predictions for microscopic phenomena which could be experimentally
distinguished from the predictions of standard quantum theory. 
This  possibility
suggests several questions.  For example, if one
performs an experiment using a real (hence not infinitely-accurate)
clock  and hopes
to observe evidence of energy decoherence, one might ask whether 
the existence of a 
more-nearly-accurate clock somewhere else in the universe could affect the 
decoherence one was hoping to detect.  
Not surprisingly, the answer to this question is
``no'' (and I do not mean to imply that any of the referenced authors
have ever suggested otherwise), but in contemplating similar
questions it might be helpful to  consider a
simple story~\cite{story} in which standard  non-relativistic quantum theory
is supposed to hold
exactly, and in which it is assumed that there is an
arbitrarily-accurate clock, but which will nevertheless lead to energy
decoherence of  the type  mentioned above.
   
Let us then imagine that two students whose names  are Alice and Bob
are taking an 
examination for their
class in quantum mechanics.  They are told that there is an isolated
quantum system  $\cal Q$ whose time-independent Hamiltonian is
$H_{\cal Q}$, and which at time $t=0$ is in the state $\rho _{\cal
  Q}(0)$.
They are asked what is the state of that system ``now'' (i.\ e., at
the time the question is asked).

There is a clock on the wall of the examination room, which for the
purpose of this story I take to be
infinitely accurate and capable of being read with
infinite precision.  Alice looks at this clock, and sees that it reads a
time I denote as $t_A$; she sets $\hbar = 1$, and writes as her answer 
\begin{equation}  \rho _{{\cal Q},A}(t_{A}) = 
                  e^{-iH_{{\cal Q}}t_{A}} \rho _{\cal Q}(0)  
                 e^{iH_{{\cal Q}} t_{A}}.          
\label{rqa1}\end{equation}

Bob does not look at the clock on the wall; instead, he looks at his
inexpensive 
wristwatch, which reads a time I denote as $t_B$.  Since he knows that
his wristwatch is not accurate, he does not assume that the actual
time (i.e., the time shown on the wall clock) is equal to $t_B$; from his
experience with the watch, he assigns a probability density I denote
as $P(t|t_{B})$ for the actual time to be $t$ given that his watch
reads $t_B$. He has learned in class that, 
if the time were surely $t$, the state of
the system $\cal Q$ would be 
\begin{equation}  \rho _{{\cal Q}}(t) = 
                  e^{-iH_{{\cal Q}}t} \rho _{\cal Q}(0)  
                 e^{iH_{{\cal Q}} t};          
\label{rq1}\end{equation}
since he thinks that the time is $t$ with probability $P(t|t_{B})$,
his answer to the examination question is~\cite{Grade}
\begin{equation} \rho _{{\cal Q},B}(t_{B}) = 
                  \int dt\, P(t|t_{B})  e^{-iH_{{\cal Q}}t} \rho _{\cal
		    Q}(0)  
                  e^{iH_{{\cal Q}}t}.
\label{rqb1}\end{equation} From equations~\ref{rqa1} and~\ref{rqb1}, 
we can see that
\begin{equation} \rho _{{\cal Q},B}(t_{B}) =
              \int dt\, P(t|t_{B}) \rho _{{\cal Q},A}(t).
\label{rqb2}\end{equation}

It has been pointed out by Poulin~\cite{DP} that this equation 
arises from relative-time calculations.  Also, a special case of this
equation arises in energy-driven collapse theories.  To see that,   
take $P(t|t_{B})$ to be a Gaussian in $(t_{B}-t)$ with width
proportional to   $\sqrt{t_{B}}$, that is, take
\begin{equation}
 P(t|t_{B}) = (2\pi \lambda
t_{B})^{-\frac{1}{2}}\exp (-(t_{B}-t)^2/(2\lambda t_{B}));
\label{ptt}\end{equation}
then eq.~\ref{rqb2} can be re-written, with the variable 
$\eta = (t_{B} -t)/\sqrt{\lambda t_{B}}$, as
\begin{equation} \rho _{{\cal Q},B}(t_{B}) = (2\pi )^{-\frac{1}{2}}
      \int d\eta \, e^{-\eta ^{2}/2}\rho _{{\cal Q},A}(t_{B}-(\lambda
      t_{B})^{\frac{1}{2}}\eta ).
\label{rqb3}\end{equation}
This is the same expression for the quantum state as Pearle~\cite{PR}
has shown to arise  in collapse
theories.  Thus properties of $  \rho _{{\cal Q},B}$     will also be 
valid in either relative-time or
energy-driven collapse theories.    This is true even though in the
story of the quantum-mechanics exam 
I had stipulated that quantum states evolve unitarily (with no modification
for collapse) and that there does exist an arbitrarily-accurate
clock; Bob just does not happen to have looked at it.

Here are some properties of the states assigned by Alice and Bob:
\begin{itemize}
\item If $P(t|t_{B})$ were given by $\delta (t-t_{B}) $ (that is, 
  if Bob's watch were in
  fact completely accurate) then it would follow from eq.~\ref{rqb2} that 
  $\rho _{{\cal Q},B}=
  \rho _{{\cal Q},A}$.  Also, $\rho _{{\cal Q},B}=
  \rho _{{\cal Q},A}$ if  $ \rho _{{\cal Q},A}(t)$ is constant for
  those values of $t$ for which $P(t|t_{B}) \neq 0$. (That is, it  would
  not matter that Bob does not know what time it is if $ \rho
  _{{\cal Q},A}$ did not depend on time.)  But in general,  $\rho
  _{{\cal Q},B} \neq  \rho _{{\cal Q},A}$.
\item Under unitary time evolution a pure state remains pure.  If
  $\rho _{\cal Q}(0)$ is a pure state, then $ \rho _{{\cal Q},A}$ will
  be pure also,  but  $\rho _{{\cal Q},B}$ need not be
  pure.  In fact, in the case in which $ \rho _{{\cal Q},A}$ is pure,  
  eq.~\ref{rqb2} gives an
  ensemble decomposition for $\rho _{{\cal Q},B}$.  According to Bob,
  the initially-pure state $\rho _{\cal Q}(0)$ has evolved into the
  mixed state $\rho _{{\cal Q},B}$.
\item Consider matrix elements of $\rho$ in a basis of energy
  eigenstates.  From eq.~\ref{rqa1},
  \begin{equation} [\rho _{{\cal Q},A}(t_A)]_{i,j} =
                   \exp (i(E_{j}-E_{i})t_{A})[\rho _{{\cal Q}}(0)]_{i,j}
  \label{rqa2}\end{equation}
  If $E_{i}=E_{j}$ (and {\it a fortiori} if $i=j$) then 
  $[\rho _{{\cal Q},A}(t_A)]_{i,j}$ is independent of $t_A$, and so
  from eq.~\ref{rqb2} $ [\rho _{{\cal Q},B}(t_B)]_{i,j}$ is also
  independent 
  of $t_B$; in fact in this case $ [\rho _{{\cal Q},B}]_{i,j}=
  [\rho _{{\cal Q},A}]_{i,j}$.  But if $E_{i}\neq E_{j}$,
  eqs.~\ref{rqb2} and~\ref{rqa2}
  show that $| [\rho _{{\cal Q},B}(t_B)]_{i,j}| 
  < |[\rho _{{\cal Q},A}]_{i,j}|$; this implies that there is 
  decoherence in energy.
  And if $P(t|t_{B})$ is sufficiently broad as a function of $t$, then
  $ [\rho _{{\cal Q},B}(t_B)]_{i,j} \rightarrow 0$, in which case 
  $\rho _{{\cal Q},B}(t_B)$ becomes independent of $t_B$ (even if
  $\rho _{{\cal Q},A}(t_A)$ is not independent of $t_A$); I will call
  this ``complete'' energy decoherence.

  Now suppose that $N$ represents an observable of the
  quantum system ${\cal Q}$, and that Alice and Bob are each asked to
  calculate the expected value of $N$.  Alice will calculate
  $\langle N\rangle _{A} = Tr[\hat{N}\rho _{{\cal Q},A}]$, and 
  Bob will calculate
  $\langle N\rangle _{B} = Tr[\hat{N}\rho _{{\cal Q},B}]$.
  In the case of complete energy
  decoherence, $\langle N\rangle _B$ will be independent of $t_B$, 
  although $\langle N\rangle _A$
  need not be independent of $t_A$. 
\item Suppose that system $\cal Q$ is composed of two subsystems $\cal
  S$ and $\cal C$ which are dynamically independent, i.e.\ that 
  $H_{\cal Q} = H_{\cal S}+ H_{\cal C}$; suppose also that these two
  subsystems are initially uncorrelated, i.e.\ that $\rho _{\cal Q}(0) =
  \rho _{\cal S}(0)\otimes \rho _{\cal C}(0)$.  Then $\rho _{{\cal
  Q},A}(t_{A}$) will also be a product: $\rho _{{\cal Q},A}(t_{A})=
  \rho _{{\cal S},A}(t_{A})\otimes \rho _{{\cal C},A}(t_{A})$,  but
  $\rho _{{\cal Q},B}(t_{B}$) will in general not be a product~\cite{Sep}.
  If Bob were asked to write the state of
  $\cal S$  he could either, in analogy to eq.~\ref{rqb1}, write
\begin{equation} \rho _{{\cal S},B}(t_{B}) = 
                  \int dt\,  P(t|t_{B})  e^{-iH_{{\cal S}}t} 
                  \rho _{\cal S}(0)  e^{iH_{{\cal S}}t},
\label{rqb4}\end{equation}  
or he could calculate
\begin{equation}\rho _{{\cal S},B}(t_{B})=Tr_{\cal C}
        [\rho _{{\cal Q},B}(t_{B})]
\label{rqb5}\end{equation}
with $\rho _{{\cal Q},B}(t_{B})$ given by eq.~\ref{rqb1}; 
he would get the same
answer either way.  Similarly, Bob could calculate 
$\rho _{{\cal C},B}(t_{B})=Tr_{\cal S}[\rho _{{\cal Q},B}(t_{B})]$, but
$\rho _{{\cal Q},B}$ would {\em not}
in general be equal to the tensor product of 
$\rho _{{\cal S},B}$ and $\rho _{{\cal C},B}$. 
\item Now suppose that $N$ is an observable of the subsystem ${\cal
  S}$, and that subsystem ${\cal C}$ is itself a clock (which 
  will be assumed to agree arbitrarily well with the clock on the
  wall). Let $T$ be an observable of  ${\cal C}$, with orthonormal
  eigenvectors $\{ |t\rangle \} $, and take~\cite{cont}
\begin{equation}
\rho _{{\cal C},A}(t_{A}) = |t_{A}\rangle \langle t_{A}|. 
\label{rc1}\end{equation}
Alice and Bob are  each asked the following question: ``If $N$ and
$T$ were now measured and the value of $T$ were found to be $t$, what
would you expect for the value of $N$?''

Alice knows that the value $t$ agrees with $t_A$, so she could simply
write 
\begin{equation}
\langle N \rangle _{A} = Tr[\hat{N}\rho_{{\cal S},A}(t_{A})].
\label{na1}\end{equation}
Alternatively, she could calculate (letting $[\langle N\rangle |t]$ denote
the expectation value of $N$ given that $t$ had been found)
\begin{equation}
[\langle N \rangle |t]_{A} = \frac{Tr[\hat{N}P_{t}\rho_{{\cal
      Q},A}(t_{A})]}{Tr[P_{t}\rho_{{\cal Q},A}(t_{A})]}
\label{na2}\end{equation}
where $\rho_{{\cal Q},A}(t_{A}) = \rho_{{\cal S},A}(t_{A})\otimes
|t_{A}\rangle \langle t_{A}|$, and $P_t$ projects ${\cal C}$
onto $|t\rangle $; of course she would get the same
answer either way.

Bob can calculate
\begin{equation}
[\langle N \rangle |t]_{B} = \frac{Tr[\hat{N}P_{t}\rho_{{\cal
      Q},B}(t_{B})]}{Tr[P_{t}\rho_{{\cal Q},B}(t_{B})]}
\label{nb1}\end{equation}
where 
\begin{equation}
\rho_{{\cal Q},B}(t_{B}) = 
\int dt_{A} P(t_{A}|t_{B})\rho_{{\cal S},A}(t_{A})\otimes
|t_{A}\rangle \langle t_{A}|,
\label{rqb6}\end{equation}
and since $Tr[\rho _{{\cal S},A}\otimes |t\rangle \langle t|] =
1$ and $Tr[\hat{N} \rho _{{\cal S},A}\otimes |t\rangle \langle t|] = 
        Tr[\hat{N} \rho _{{\cal S},A}]$, Bob will find that
\begin{equation}
[\langle N \rangle |t]_{B} = Tr[\hat{N}\rho_{{\cal S},A}(t)].
\label{nb2}\end{equation}
Thus even though $\langle N\rangle_B$ is in general different than
$\langle N\rangle_A$ (in fact, in the case of complete energy decoherence
$\langle N\rangle_B$ is independent of $t_B$ while $\langle
N\rangle_A$ need not be independent of $t_A$), Bob gets the same
expression for $[\langle N\rangle |t]$ as does Alice.  The reason that 
$\langle N\rangle_B$ and $\langle N\rangle_A$ differ is that Alice
knows what time it is and Bob does not.  However, since the reading of
the internal clock ${\cal C}$ and the wall clock agree, once Bob
learns the value found for the observable $T$ he does know what
Alice knows, and so his expectation for $N$ is the same as Alice's.
For Alice, ${\cal S}$ and ${\cal C}$ are not correlated; for Bob they
are, and by using eq.~\ref{nb1} he exploits this correlation (as represented
in $\rho _{{\cal Q},B}$) to reproduce Alice's result: $[\langle
N\rangle |t]_B$ (in eq.~\ref{nb2}) is the same function of $t$
as $\langle N\rangle _A$ (in eq.~\ref{na1}) is of $t_A$.

\item Let's again think of ${\cal Q}$ as a single system, and let $N$ be
  an observable of ${\cal Q}$. Say Bob is asked the following
  question: ``If you were to look at the clock on the wall and see a
  time $t$, what would you expect to find for the value of $N$?'' He
  could of course just do what Alice does; he knows that if the clock
  on the wall reads $t$ then the state of ${\cal Q}$ is as given in 
  eq.~\ref{rq1}, and that $\langle N\rangle = Tr[\hat{N}\rho_{{\cal Q}}]$. 
  However, suppose for some reason he in enamored of the procedure
  described previously, which is to consult his watch rather than the wall
  clock to write a quantum state, and then to find the
  expectation of $N$ from that quantum state.  He could consider the
  compound system consisting of ${\cal Q}$ together with the wall
  clock, in which case the wall clock would be an internal clock of
  that compound system, so he could proceed as above: write the state
  of the compound system as in eq.~\ref{rqb1}, and then find the
  expectation of $N$ given that the wall clock reads $t$, as in
  eq.~\ref{nb1} (with the compound system replacing ${\cal Q}$ in
  those equations).  Of course he would get the same answer either
  way, just as his result (eq.~\ref{nb2}) agrees with Alice's 
  result (eq.~\ref{na1}).
\end{itemize}

The moral of this story is that although
$\rho _{{\cal Q},B}(t_{B})$ 
exhibits energy decoherence (and in the extreme case of complete energy
decoherence is independent of $t_B$),  nevertheless if ${\cal Q}$
contains a subsystem which is an accurate clock (or, which amounts to
the same thing, if  ${\cal Q}$ can be expanded so as to include such a
subsystem) then Bob can recover the time dependence of a quantum
system as seen by Alice by considering the correlation between that
system and the internal clock.  This story might help us to interpret
energy-driven collapse theories.  Those theories postulate that
unitary evolution of quantum states is supplemented by collapse
onto states of definite total energy~\cite{TE}.  As
shown in eq.~5.6a of Pearle~\cite{PR}, in these theories the
state of a quantum system ${\cal Q}$ at time $t$ is given by 
\begin{equation}
\rho _{\cal Q}(t) = (2\pi )^{-\frac{1}{2}}\int d\eta\, e^{-\eta ^{2}/2}
   (\rho_{\cal Q})^{U}(t-(\lambda t)^{\frac{1}{2}}\eta ),
\label{rq2}\end{equation}
where $(\rho_{\cal Q})^{U}$ represents what the state would be if it
did evolve unitarily, that is
\begin{equation}
(\rho_{\cal Q})^{U}(t) := e^{-iH_{\cal Q}t}\rho _{\cal Q}(0)e^{iH_{\cal Q}t}.
\label{rq3}\end{equation}
The theory of ref.~\cite{Mil}, also, leads to an expression for $\rho
_{\cal Q}$ which is essentially identical to that in eq.~\ref{rq2}.
    
The relevance of the story of the quantum-mechanics exam is, of
course, that the
expression for $\rho _{{\cal Q},B}$ given in eq.~\ref{rqb3} is the
same as that for $\rho _{\cal Q}$ given in eq.~\ref{rq2}. So now we
can ask what  an energy-driven collapse theory would imply for the
time-dependence of the expectation
value of some observable $N$ of some quantum system ${\cal S}$.  One
way to answer this question would be to apply eq.~\ref{rq2} to  
the  compound system   
${\cal Q}$ consisting of  ${\cal S}$ together with a clock.  The clock
would then be an internal clock; just as Bob, in spite of the energy
decoherence of $\rho _{{\cal Q},B}$ can reproduce the
time-dependence as seen by Alice by considering the correlation
between ${\cal S}$ and the clock, so too would the time dependence of
$\langle N\rangle $ (with respect to that clock) as predicted by an
energy-driven collapse theory be exactly the same
as that predicted by standard quantum theory, i.e. without
energy-driven collapse.  This result has been shown in some detail by
Simon and Jaksch~\cite{SJ}; see also ref.~\cite{JF} for a similar
result for the theory of ref.~\cite{Mil}.

On the other hand, Pearle~\cite{PR} has suggested a second way in which
one might calculate the time dependence of $\langle N\rangle $ in an
energy-driven collapse theory: take ${\cal Q}$ in eq.~\ref{rq2}
to be just ${\cal S}$ by itself, and say as usual that 
$\langle N\rangle = Tr[\hat{N}\rho _{\cal Q}(t)]$.  But then, as
pointed out by Pearle, there seems to be a paradox: although according
to the first
method of calculation $\langle N\rangle $ agrees with
standard quantum theory, according to the second it does not; for
example, in the case of complete energy decoherence, the second method
would imply that $\langle N\rangle $ was constant even if the first
implied it was not constant.

Since any clock could be considered to be a quantum system, it is
difficult to see how the first method of calculation, which applies
the evolution predicted by an energy-driven collapse theory
(eq.\ref{rq2}) to the combined system of ${\cal S}$ and the clock,
could be incorrect within that theory.  Because within such a theory
the value of $t$ is not directly accessible (since clocks as well as
anything else are supposed to suffer energy-driven collapse), one
might perhaps deny the validity of the second method of calculation.
One would thereby avoid Pearle's paradox, but at the cost of allowing
that the theory is observationally completely equivalent to standard
quantum theory.  It seems then that these theories, if interpreted to be 
non-paradoxical, not only do not lead to testable predictions for
microscopic phenomena which differ from those of standard quantum
theory but also do not help to reconcile quantum theory
with the observed macroscopic world. 

Whether or not there is energy-driven collapse, as long as one has a
sufficiently-accurate clock one should expect that results of
experiments on microscopic systems would agree with predictions of  
standard quantum theory.  But since, as stressed in
refs~\cite{SW}--\cite{GPP}, real clocks are not infinitely accurate,
this agreement would not be exact.  
This is illustrated in the story of the quantum-mechanics exam  by the
fact that $\rho _{{\cal Q},B}$ does not coincide with  $\rho _{{\cal Q},A}$;
if Bob does not use an accurate clock his expectation for measurements on
${\cal Q}$ will not be the same as Alice's.  Note that the mere {\em
existence} of an accurate clock does not affect Bob's analysis.  Bob
could have written eq.~\ref{rqb1} even if there were in fact no clock
on the wall; the clock on the wall makes no difference to Bob if he does not
look at it.

More generally, any experiment will involve apparatus---clocks,
lasers, whatever---which necessarily have some inaccuracy.  And whether
this inaccuracy comes from limits set by quantum gravity (as
suggested in ref~\cite{GPP}) or from the more mundane fact that the
apparatus may be cheaply manufactured, it can certainly influence the
experimental results.  In any experiment, it is important to
understand the accuracy of the apparatus which is actually used;
the possibility of more-nearly-accurate apparatus somewhere else in
the universe is not relevant.     

{\bf Added note:} After the original version of this paper was
posted, I noticed an (earlier)
paper by Bartlett, Rudolph and Spekkens (BRS)~\cite{BRS}, and I realized
that the story of the quantum-mechanics exam  can also serve as an
illustration of points made in that paper.   BRS
argue that whether or not coherences (i.\ e.\ off-diagonal elements
of $\rho$) vanish can depend on the way the system is described.  That
is what does happen with Alice and Bob; in the case I have called
complete energy decoherence (which is the case in which Bob's
wristwatch is no good at all) matrix elements of $\rho_{{\cal Q},B}$
between states of different energy vanish, while those of
$\rho_{{\cal Q},A}$ might not. Thus Bob plays the role of the
``fictionist'', and Alice the role of the ``factist'' mentioned in the
title of BRS.
BRS argue further that
the choice of description would not lead to different predictions for
experiments.  Since neither Alice nor Bob has made any mistake, they
certainly would not give contradictory predictions, and in the case in
which ${\cal Q}$ has a subsystem which is an accurate clock,
eqs.~\ref{na1} and \ref{nb2} show that their predictions would be the
same.  In the terminology used by BRS, the clock-subsystem of
${\cal Q}$ serves as an ``internal reference frame'', the clock on the wall
as an ``implicated external reference frame'', and 
Bob's watch, if sufficiently inaccurate, as a ``nonimplicated external
reference frame''.

\vspace{.6cm}
Acknowledgments: I thank Philip Pearle for an enlightening correspondence 
about energy-driven collapse theories.  I acknowledge also 
the hospitality of the Lawrence Berkeley National Laboratory.


\begin{thebibliography}{99}

\bibitem{SW} H.\ Salecker and E.\ P.\ Wigner, Phys.\ Rev.\  {\bf 109},
             571 (1958).
\bibitem{AP} A.\ Peres, Am.\ J.\ Phys.\ {\bf 48}, 552 (1980).
\bibitem{PW} D.\ N.\ Page and W.\ K.\ Wooters, Phys.\ Rev.\ D {\bf 27},
             2885 (1983).
\bibitem{EGR} I.\ L.\ Egusquiza, L.\ J.\ Garay, and J.\ M.\ Raya,
              Phys.\ Rev. A {\bf 59}, 3236 (1999).
\bibitem{GPP} R.\ Gambini, R.\ A.\ Porto, and J.\ Pullin, Class.\
              Quant.\ Grav.\ {\bf 21}, L51 (2004).
\bibitem{Mil} G.\ J.\ Milburn, Phys.\ Rev.\ A {\bf 44}, 5401 (1991).
\bibitem{BW} D.\ Bedford and D.\ Wang, Nuovo Cimento B {\bf 37}, 55 (1977). 
\bibitem{Hu} L.\ P.\ Hughston, Proc.\ R.\ Soc. London A {\bf 452}, 953 (1996).
\bibitem{AH} S.\ L.\ Adler and L.\ P.\ Horwitz, J.\ Math.\ Phys.\ {\bf
             41}, 2485 (2000).
\bibitem{story} Story-telling has been  said to be a useful
          pedagogical technique for mathematics, so why not try it here? 
          See S.\ Tomlin, Nature {\bf 436}, 622 (2005).
\bibitem{Grade} The grade that Bob receives on the examination may
  depend on the foundational views of the person doing the grading.
  Perhaps that person holds the view that the state Bob assigns
  to a quantum system should just represent the knowledge that Bob has about
  the system; then Bob would surely receive the grade of A, since
  eq.~\ref{rqb1} does correctly represent what he knows about ${\cal
  Q}$. On the other hand, perhaps the grader holds the view that the
  state of ${\cal Q}$ {\em really is} that given in eq.~\ref{rqa1}, 
  and therefore
  that Bob, because he has not looked at the clock, is ignorant of
  what  that state really is.  According to this view, Bob's
  answer (eq.~\ref{rqb1}) would need the ``ignorance interpretation'',
  but this should be worth at least a grade of B.
\bibitem{DP} D.\ Poulin, ``Toy Model for a Relational Formulation of
                          Quantum Theory'', preprint quant-ph/0505081 (2005).
\bibitem{PR} P.\ Pearle, Phys.\ Rev.\ A {\bf 69}, 042106 (2004).
\bibitem{Sep} Under the assumption that $\rho _{\cal Q}(0)$, and hence
       $\rho _{{\cal Q},A}(t_{A})$, is
       a   tensor product, eq.~\ref{rqb2} shows that $\rho _{{\cal
         Q},B}(t_{B})$ is separable, not entangled (see for example
        A.\ Sen(De), U.\ Sen, M.\ Lewenstein, and A. Sanpera, 
         ``Lectures on Quantum Information Chapter 1: The separability
         versus entanglement problem'', preprint quant-ph/0508032
         (2005)); nevertheless, for Bob ${\cal S}$ and ${\cal C}$ are
         (classically) correlated.
\bibitem{cont} For simplicity, I will pretend that the spectrum of
         $\hat{T}$ is discrete, thus avoiding the need for wave packets.
\bibitem{TE} Comments below do not apply to other theories, in which
            the collapse in onto states of definite {\em local} energy.
\bibitem{SJ} C.\ Simon and D.\ Jakish, Phys.\ Rev.\ A {\bf 70}, 052104
          (2004).
\bibitem{JF} J.\ Finkelstein, Phys.\ Rev. A {\bf 47}, 2412 (1993).
\bibitem{BRS} S.\ D.\ Bartlett, T.\ Rudolph, and R.\ W.\ Spekkens,
              ``Dialogue Concerning Two Views on Quantum Coherence:
              Factist and Fictionist'', preprint quant-ph/0507214 (2005).
\end{thebibliography}
\end{document}